\documentclass[conference]{IEEEtran}

\usepackage{times,latexsym,amsfonts,amsmath,amssymb,mathrsfs,verbatim,cite}
\usepackage{epsfig,epsf}
\usepackage{graphicx}
\usepackage[dvips]{color}
\usepackage{txfonts}

\def\nN{{\mathbb N}}
\def\zZ{{\mathbb Z}}
\def\rR{{\mathbb R}}
\def\eE{{\mathbb E}}
\def\pP{{\mathbb P}}

\def\QED{\mbox{\rule[0pt]{1.5ex}{1.5ex}}}

\def\@begintheorem#1#2{\tmpitemindent\itemindent\topsep 0pt\rm\trivlist
    \item[\hskip \labelsep{\indent\it #1\ #2:}]\itemindent\tmpitemindent}
\def\@opargbegintheorem#1#2#3{\tmpitemindent\itemindent\topsep 0pt\rm \trivlist
    \item[\hskip\labelsep{\indent\it #1\ #2\
    \rm(#3):}]\itemindent\tmpitemindent}
\def\@endtheorem{\endtrivlist\unskip}

\newtheorem{theorem}{Theorem}
\newtheorem{definition}{Definition}
\newtheorem{fact}{Fact}
\newtheorem{proposition}{Proposition}
\newtheorem{lemma}{Lemma}
\newtheorem{corollary}{Corollary}

\renewcommand{\theequation}{\arabic{section}.\arabic{equation}}

\newcommand{\supp}{\operatorname{supp}}
\begin{document}

\title{Information-Theoretic Bounds for Multiround Function Computation in Collocated Networks$^{\text{\small 1}}$}



\author{\IEEEauthorblockN{Nan Ma}
\IEEEauthorblockA{ECE Dept, Boston University\\
Boston, MA 02215 \\ {\tt nanma@bu.edu}}
\and
\IEEEauthorblockN{Prakash Ishwar}
\IEEEauthorblockA{ECE Dept, Boston University\\
 Boston, MA 02215 \\ {\tt pi@bu.edu}}
\and
\IEEEauthorblockN{Piyush Gupta}
\IEEEauthorblockA{Bell labs, Alcatel-Lucent \\
Murray Hill, NJ 07974\\ {\tt pgupta@research.bell-labs.com}}
}

\maketitle

\begin{abstract}
We study the limits of communication efficiency for function
computation in collocated networks within the framework of
multi-terminal block source coding theory.  With the goal of computing
a desired function of sources at a sink, nodes interact with each
other through a sequence of error-free, network-wide broadcasts of
finite-rate messages. For any function of independent sources, we
derive a computable characterization of the set of all feasible
message coding rates - the rate region - in terms of single-letter
information measures.  We show that when computing symmetric functions
of binary sources, the sink will inevitably learn certain additional
information which is not demanded in computing the function. This
conceptual understanding leads to new improved bounds for the minimum
sum-rate. The new bounds are shown to be orderwise better than those
based on cut-sets as the network scales. The scaling law of the
minimum sum-rate is explored for different classes of symmetric
functions and source parameters.
\end{abstract}

\section{Introduction}
\addtocounter{footnote}{+1} \footnotetext{The work of N.~Ma and
P.~Ishwar was supported by the US National Science Foundation (NSF)
under award (CAREER) CCF--0546598. The work of P.~Gupta was
supported in part by NSF Grant CNS-0519535. Any opinions, findings,
and conclusions or recommendations expressed in this material are
those of the authors and do not necessarily reflect the views of the
NSF.  }

Both wired and wireless data networks such as the Internet and the
mobile ad hoc and wireless mesh networks have been designed with the
goal of efficient data transfer as opposed to data processing. As a
result, computation takes place only after all the relevant data is
moved. Two-way interaction would be utilized to primarily improve the
reliability of data-reproduction than data processing
efficiency. However, to maximize the data processing efficiency, it
may be necessary for nodes to interact bidirectionally in multiple
rounds to perform distributed computations in the network. In this
paper we attempt to formalize this intuition through a distributed
function computation problem where data processing efficiency is
measured in terms of the total number of bits exchanged per sample
computed. Our objective is to study the fundamental limits of
multiround function computation efficiency within a distributed source
coding framework, involving block-coding asymptotics and vanishing
probability of function-computation error, for ``collocated'' networks
where broadcasted messages can be heard by all nodes. We derive an
information-theoretic characterization of the set of feasible
coding-rates and explore the benefit of multiround communication.

This problem was studied in \cite{Kumar2005} within a communication
complexity framework where computation is required to be error-free.
For collocated networks and random planar multihop networks, the
scaling law of the maximum rate of computation with respect to a
growing size of the network was derived for divisible functions and
two subclasses of symmetric functions namely type-sensitive and
type-threshold functions. This work was extended in \cite{GuptaISIT}
to multihop networks having a finite maximum degree. In
\cite{GuptaISIT} it was also shown that for any network, if a
nonzero per-sample error probability was allowed, the computation of
a type-sensitive function could be converted to that of a
type-threshold function. In \cite{MassimoAllerton} a min-cut bound
was developed for acyclic network topology and was shown to be tight
for tree networks and divisible functions.

In \cite{Prabha}, a function computation problem in a collocated
network was posed within a distributed block source coding framework,
under the assumption that conditioned on the desired function, the
observations of source nodes are independent. An information-theoretic
lower bound for the sum-rate-distortion function was derived. It was
shown that if the desired function and the observation noises are
Gaussian, the lower bound is tight and there is no advantage to be
gained, in terms of sum-rate, by broadcasting messages, in comparison
to sending messages through point-to-point links from source nodes to
the sink where the function is desired to be computed. Multiround
(interactive) function computation in a two-terminal network was
studied in \cite{OrlitskyRoche,ISIT08} within a distributed block
source coding framework.

The impact of transmission noise on function computation was
considered in \cite{Gallager88,Srikant,Munther} but without a block
coding rate, i.e., only one source sample is available at each node.
A joint source-channel function computation problem over
noninteractive multiple-access channels was studied in
\cite{GastparMAC}. Our focus is on the block source coding aspects
of function computation and we assume that message exchanges are
error-free.

The present work studies a multiround function computation problem in
a collocated network within a multi-terminal source coding framework
described in Sec.~\ref{sec:problem}. Sensors observe discrete
memoryless stationary sources taking values in finite alphabets.  The
goal is to compute a samplewise function at a sink with a probability
which tends to one as the block-length tends to infinity. We derive a
computable characterization of the rate region and the minimum
sum-rate in terms of information quantities
(Sec.~\ref{sec:rateregion}).
For computation of symmetric functions of binary sources, the sink is
shown to inevitably obtain certain additional information, which is
not demanded in computing the function (Sec.~\ref{subsec:extra}). This
key observation is formalized under the vanishing block-error
probability criterion (Lemma~\ref{lemma:individualinterval}) and also
the zero-error criterion (Lemma~\ref{lemma:zeroerror}). This
conceptual understanding leads to improved bounds for the minimum
sum-rate (Sec.~\ref{subsec:bounds}). These bounds are shown to be
orderwise better than cut-set bounds as the size of the network
grows. The scaling law of the minimum sum-rate is evaluated in
different cases in Sec.~\ref{subsec:scaling}.

\section{Multiround Computation in Collocated Networks}
\label{sec:problem}

Consider a network consisting of $m$ source nodes numbered
$1,\ldots,m$, and one (un-numbered) sink (node).  Each source node
observes a discrete memoryless stationary source taking values in a
finite alphabet. The sink has no source samples. For each $j \in
[1,m]$,\footnote{When $a$ and $b$ are integers, $[a,b]$ denotes an
integer interval, which is the set of all consecutive integers
beginning with $a$ and ending with $b$.}  let $\mathbf
X_j:=(X_j(1),\ldots,X_j(n))\in (\mathcal{X}_j)^n$ denote the $n$
source samples which are available at node-$j$. To isolate the impact
of the structure of the desired function on the efficiency of
computation, we assume sources are independent, i.e., for
$i=1,\ldots,n$, $(X_1(i),X_2(i),\ldots,X_m(i))\sim$ iid $p_{X^m} =
\prod_{j=1}^m p_{X_j}$. Let $f:
\mathcal{X}_1\times\ldots\times\mathcal{X}_m \rightarrow \mathcal{Z}$
be the function of interest at the sink and let
$Z(i):=f(X_1(i),\ldots,X_m(i))$. The tuple $\mathbf
Z:=(Z(1),\ldots,Z(n))$, which denotes $n$ samples of the samplewise
function of all the sources, is desired to be computed at the sink.

The communication takes place over $r$ rounds. In each round, source
nodes broadcast messages according to the schedule $1,\ldots,m$.  Each
message depends on the source samples and all the previous messages
which are available to the broadcasting node. Nodes are collocated,
meaning that every broadcasted message is recovered without error at
every node. After $mr$ message broadcasts over $r$ rounds, the sink
computes the samplewise function based on all the messages.

\begin{definition}\label{def:code}
An $r$-round distributed block source code for function computation in
a collocated network with parameters $(r,n,|{\mathcal
M}_1|,\ldots,|{\mathcal M}_t|)$ is the tuple $(e_1,\ldots,e_t,g)$ of
$t:= mr$ block encoding functions $e_1,\ldots,e_t$ and a block
decoding functions $g$, of block-length $n$, where for every $j \in
[1,t]$, $k=(j \!\! \mod m)$,\footnote{$k=(j \!\! \mod m)$ means that
$k \in [1,m]$ and $m$ divides $(k-j)$.}
\begin{equation*}
e_j: \left(\mathcal X_k\right)^{n} \times \bigotimes_{i=1}^{j-1}
{\mathcal M}_i \rightarrow {\mathcal M}_j,\ \ \  g:
\bigotimes_{j=1}^{t} {\mathcal M}_j \rightarrow {\mathcal Z}.
\end{equation*}
The output of $e_j$, denoted by $M_j$, is called the $j$-th message,
$r$ is the number of rounds, and $t$ is the total number of
messages. The output of $g$ is denoted by $\widehat{\mathbf Z}$. For
each $j$, $(1/n) \log_2 |{\mathcal M}_j|$ is called the $j$-th
block-coding rate (in bits per sample).
\end{definition}

\emph{Remarks:} (i) Each message $M_j$ could be a null message
($|\mathcal{M}_j|=1$). By incorporating null messages, the multiround
coding scheme described above subsumes all orders of messages
transfers from $m$ source nodes, and an $r$-round coding scheme
subsumes an $r'$-round coding scheme if $r'<r$. (ii) Since the
information available to the sink is also available to all source
nodes, there is no advantage in terms of sum-rate to allow the sink to
send any message.

\begin{definition}\label{def:rateregion}
A rate tuple ${\mathbf R} = (R_1, \ldots, R_t)$ is admissible for
$r$-round function computation if, $\forall \epsilon > 0$, $\exists~
\bar n(\epsilon,t)$ such that $\forall n> \bar n(\epsilon,t)$, there
exists an $r$-round distributed block source code with parameters
$(r,n,|{\mathcal M}_1|,\ldots,|{\mathcal M}_t|)$ satisfying
\begin{equation*}
\forall j \in [1, t],\ \frac{1}{n}\log_2 |{\mathcal M}_j| \leq R_j +
\epsilon, \ \ \pP(\widehat{\mathbf Z} \neq {\mathbf Z}) \leq \epsilon.
\end{equation*}
\end{definition}

The set of all admissible rate tuples, denoted by ${\mathcal R}_r$, is
called the operational rate region for $r$-round function
computation. The minimum sum-rate $R_{sum,r}$ is given by
$\min_{{\mathbf R} \in {\mathcal R}_r} \left(\sum_{j=1}^t
R_j\right)$. Note that since each message could be a null message, if
$r'<r$, $R_{sum,r'}\geq R_{sum,r}$ holds. The goal of this work is to
obtain a single-letter characterization of the rate region (a
computable characterization independent of block-length $n$), to study
the scaling behavior of $R_{sum,r}$, and to investigate the benefit of
multiround function computation.

\section{Rate Region}\label{sec:rateregion}

The rate region for $r$-round function computation for $m$ independent
sources can be characterized by Theorem~\ref{thm:rateregion}, in terms
of single-letter mutual information quantities involving auxiliary
random variables satisfying Markov chain and conditional entropy
constraints.

\begin{theorem}\label{thm:rateregion}
\begin{eqnarray}
\lefteqn{ \mathcal R_r = \{\mathbf R~|~\exists \ U^t,
\mbox{s.t. } \forall j \in [1,t], \mbox{ and } k=(j \!\!\! \mod m),} && \nonumber\\
&& R_j \geq I(X_k;U_j|U^{j-1}), U_j-(U^{j-1},X_k)-(X^{k-1},X_{k+1}^m),\nonumber\\
 && H(f(X^m)|U^t)=0~\},\label{eqn:rateregion}
\end{eqnarray}
where $U^t$ are auxiliary random variables taking value in finite
alphabets. Cardinality bounds on the alphabets of the auxiliary
random variables can be derived using the Carath\'{e}odory theorem
but are omitted.
\end{theorem}

The proof of achievability follows from standard random conditional
coding arguments and is briefly outlined as follows. For the $j$-th
message, $j = 1, \ldots, t$, node-$k$ ($k= j\!\!  \mod m$) quantizes
$\mathbf{X}_k$ into $\mathbf{U}_{j}$ with $\mathbf{U}^{j-1}$ as side
information, which is available at every node, so that every node
can reproduce $\mathbf{U}_{j}$. After all the message transfers, the
sink produces $\mathbf{\widehat Z}$ based on $\mathbf{U}^t$. The
constraints in (\ref{eqn:rateregion}) ensure that
$\pP(\mathbf{\widehat Z}=\mathbf{Z})\rightarrow 1$ as $n\rightarrow
\infty$.

The (weak) converse, given in Appendix~\ref{app:converse}, is proved
using standard information inequalities, suitably defining auxiliary
random variables, and using time-sharing arguments. Specifically,
$U_1:=(Q,U_1(Q))$, $Q\sim$ Uniform$[1,n]$ independent of $\mathbf
X^m$, for all $q \in [1,n]$,
$U_1(q)=\{M_1,X^m(1),\ldots,X^m(q-1)\}$, and for all $j \in [2,t]$,
$U_{j}:=M_j$.

By adding all the rate inequalities in (\ref{eqn:rateregion}) and
enforcing all the constraints, we have the following characterization
of the minimum sum-rate.

\begin{corollary}\label{cor:sumrate}
\begin{eqnarray}
R_{sum,r} &=& \min_{U^t} I(X^m;U^t), \label{eqn:minsumrate}
\end{eqnarray}
where $U^t$ are subject to all the Markov chain and conditional
entropy constraints in (\ref{eqn:rateregion}).
\end{corollary}

The Markov chain and conditional entropy constraints of
(\ref{eqn:rateregion}) imply a key structural property which $U^t$
need to satisfy. This property is described below in
Lemma~\ref{lem:rectangle}.  This lemma provides a bridge between
certain fundamental concepts which have been studied in the
communication complexity literature\cite{CommComplexity} and
distributed source coding theory.  In order to state the lemma, we
need to introduce some terminology used in the communication
complexity literature\cite{CommComplexity}. A subset $\mathcal A
\subseteq \bigotimes_{i=1}^{m}\mathcal X_i$ is called a
\emph{rectangle} if for every $i \in [1,m]$, there exists $\mathcal
S_i\subseteq \mathcal X_i$ such that $\mathcal A=
\bigotimes_{i=1}^{m}\mathcal S_i$. A set $\mathcal A$ is called
\emph{$f$-monochromatic} if the function $f$ is constant on $\mathcal
A$. The support-set of a probability mass function $p$ is the set over
which it is strictly positive and is denoted by $\supp(p)$.

\begin{lemma}\label{lem:rectangle}
Let $U^t$ be any set of auxiliary random variables satisfying the
Markov chain and conditional entropy constraints in
(\ref{eqn:rateregion}). If $\supp(p_{X^m})=
\bigotimes_{i=1}^{m}\mathcal X_i$, then for any realization $u^t$ of
$U^t$, $\mathcal{A}(u^t):=\{x^m|p_{X^m U^t}(x^m,u^t)>0 \}$ is an
$f$-monochromatic rectangle in $\bigotimes_{i=1}^{m}\mathcal X_i$.
\end{lemma}

{\em Proof:} The Markov chains in (\ref{eqn:rateregion}) induce the
following factorization of the joint probability.
\begin{eqnarray*}
p_{X^m U^t}(x^m,u^t) &=& p_{X^m}(x^m) p_{U_1|X_1}(u_1|x_1)
p_{U_2|X_2U_1}(u_2|x_2,u_1)\ldots \nonumber \\
 &=:& p_{X^m}(x^m) \prod_{i=1}^m \phi_i(x_i,u^t),
\end{eqnarray*}
where $\phi_i$ is the product of all the factors having conditioning
on $x_i$. For each $i \in [1,m]$, let $\mathcal
S_i(u^t):=\{x_i~|~\phi_i(x_i,u^t)>0\}$. Since $\forall x^m$,
$p_{X^m}(x^m)>0$, we have $\mathcal A(u^t)=\bigotimes_{i=1}^m \mathcal
S_i(u^t)$. Since $H(f(X^m)|U^t)=0$ holds, $\mathcal A(u^t)$ is
$f$-monochromatic. \hspace*{\fill}~\QED


\section{Computing Symmetric Functions of Binary Sources}\label{sec:binarysymm}

In this section, we focus on the problem of computing symmetric
functions of $m$ nontrivial Bernoulli sources: $\forall i \in
[1,m]$, $\mathcal X_i=\{0,1\}$, $p_{X_i}(1)=\pi_{i}$, where
$\pi_{i}\in(0,1)$. Symmetric functions are invariant to any
permutation of their arguments. A symmetric function $f(X^m)$ of
binary sources is completely determined by the (integer) sum of the
sources $S:=\sum_{i=1}^m X_i$. In other words, $\exists
f':[0,m]\rightarrow \mathcal Z$, such that $f'(s)=f(x^m)$.

\begin{definition} Given a function $f':[0,m]\rightarrow \mathcal
Z$, an interval $[a,b]\subseteq [0,m]$ is a maximal $f'$-monochromatic
interval if (i) it is $f'$-monochromatic and (ii) it is not a proper
subset of an $f'$-monochromatic interval.
\end{definition}


The collection of all the maximal $f'$-monochromatic intervals can be
constructed as follows. First, consider all the inverse images
$\{f'^{-1}(z)\}_{z~\in \mathcal Z}:= \{\{s|s\in[0,m],
f'(s)=z\}\}_{z~\in \mathcal Z}$. Next, each inverse image can be
written as a disjoint union of nonadjacent intervals. The collection
of all such intervals from all inverse images, denoted by $\{
[a_v,b_v]\}_{v=1}^{v_{\max}}$, forms the collection of all the maximal
$f'$-monochromatic intervals. Note that they also form a partition of
$[ 0,m]$. Without loss of generality, we assume that these intervals
are ordered so that $a_1=0,b_{v_{\max}}=m$ and $\forall v \in
[2,v_{\max}], a_{v}=b_{v-1}+1$.

%

\subsection{Sink learns more than the result of function computation}
\label{subsec:extra}

Note that if $f(S) = z$ then $S \in f'^{-1}(z)$ which is, in general,
a disjoint union of several maximal $f'$-monochromatic
intervals. Thus, if the sink can successfully compute the function
$f'(S)$, one may expect that the sink can only estimate the value of
$S$ as belonging to the \emph{union} of several intervals.  Somewhat
surprisingly, however, it turns out that due to the structure of the
multiround code, the sink will, in fact, be able to identify a {\em
single} maximal monochromatic interval to which $S$ belongs as opposed
to the union of several intervals. More surprisingly, the sink will be
able to correctly identify the source-values at certain nodes.
Lemma~\ref{lemma:individualinterval} formalizes this unexpected
property and plays a central role in proving
Theorem~\ref{thm:lowersymmfuncBerp}(i).

\begin{lemma}\label{lemma:individualinterval}
Let $f(x^m)$ be a symmetric function of binary variables and
$\{[a_v,b_v]\}_{v=1}^{v_{\max}}$ the collection of all the maximal
$f'$-monochromatic intervals associated with $f$. Let $X^m$ be $m$
independent nontrivial Bernoulli random variables and $U^t$
auxiliary random variables which satisfy the Markov chain and
conditional entropy constraints in (\ref{eqn:rateregion}).  Then for
any $u^t\in \supp(p_{U^t})$, the following conditions hold.\\ (i)
There exists $v(u^t)\in [1,v_{\max}]$ such that
\[\pP(S\in[a_{v(u^t)}, b_{v(u^t)}]| U^t=u^t)=1.\]
(ii) There exist $\mathcal K_1(u^t)\subseteq [1,m]$ and $\mathcal
K_0(u^t)\subseteq[1,m]$ such that: $\mathcal K_1(u^t) \bigcap
\mathcal K_0(u^t)=\{ \}, |\mathcal K_1(u^t)|\geq a_{v(u^t)},
|\mathcal K_0(u^t)|\geq m-b_{v(u^t)}$, and
\[\pP(\forall i\in \mathcal
K_1(u^t), \forall i' \in \mathcal K_0(u^t), X_i=1,
X_{i'}=0|U^t=u^t)=1.\]
\end{lemma}
\begin{IEEEproof}
Due to Lemma~\ref{lem:rectangle}, $\mathcal{A}(u^t)=\{x^m|p_{X^m
U^t}(x^m,u^t)>0 \}$ is an $f$-monochromatic rectangle, which can be
expressed as $\bigotimes_{i=1}^m \mathcal S_i(u^t)$, where $\mathcal
S_i(u^t)$ is either $\{0\}$ or $\{1\}$ or $\{0,1\}$. Let $\mathcal
K_1(u^t):=\{i~|~ \mathcal S_i(u^t)=\{1\} \}$ and $\mathcal
K_0(u^t):=\{i~|~ \mathcal S_i(u^t)=\{0\} \}$. Let
$\alpha(u^t):=|\mathcal K_1(u^t)|$ and $\beta(u^t):= m - |\mathcal
K_0(u^t)|$. It can be shown that the projection of $\mathcal{A}(u^t)$
under the linear transformation $s=\left(\sum_{i=1}^m x_i\right)$
given by $\mathcal{A}'(u^t):=\left\{\left(\sum_{i=1}^m
x_i\right)|p_{X^m U^t}(x^m,u^t)>0 \right\}$ is an $f'$-monochromatic
interval $[\alpha(u^t),\beta(u^t)]$. Since
$\{[a_v,b_v]\}_{v=1}^{v_{\max}}$ is the collection of all the maximal
$f'$-monochromatic intervals, $\exists~v(u^t)\in [1,v_{\max}]$ such
that $\mathcal{A}'(u^t)\subseteq [a_{v(u^t)},b_{v(u^t)}]$. Therefore
(i) holds. Since $\mathcal{A}'(u^t)=[\alpha(u^t),\beta(u^t)] \subseteq
[a_{v(u^t)},b_{v(u^t)}]$, we have $\alpha(u^t)\geq a_{v(u^t)}$ and
$\beta(u^t)\leq b_{v(u^t)}$. Therefore (ii) holds.
\end{IEEEproof}

Although learning that $f'(S) = z$ is equivalent to learning that $S
\in f'^{-1}(z)$, which is generally a union of several intervals,
Lemma~\ref{lemma:individualinterval}(i) shows that the structure of
block source coding for function computation in collocated networks is
such that the sink will inevitably learn the {\em exact} interval
$[a_{v(u^t)},b_{v(u^t)}]$ in which $S$ resides even though this
information is not demanded in computing $f'(S)$. Similarly,
Lemma~\ref{lemma:individualinterval}(ii) shows that although learning
that $S\in [a_{v(u^t)},b_{v(u^t)}]$ is equivalent to learning that
there exist $a_{v(u^t)}$ nodes observing ones and $(m-b_{v(u^t)})$
nodes observing zeros, the sink will inevitably learn the identities
of these nodes.

Lemma~\ref{lemma:individualinterval} describes a property of the
single-letter characterization of the rate region. It does not, as
such, have a direct operational significance.  Hence the conclusions of
the previous paragraph can be only accepted as intuitive
interpretations.  If, however, the block-error probability criterion
$\pP(\widehat{\mathbf Z} \neq \mathbf Z)\leq \epsilon$ in
Definition~\ref{def:rateregion} is replaced by the zero-error
criterion $\pP(\widehat{\mathbf Z} \neq \mathbf Z)=0$ as in
\cite{Kumar2005}, we obtain Lemma~\ref{lemma:zeroerror} which holds
for {\em every sample realization} and provides an operational
significance to the results suggested by
Lemma~\ref{lemma:individualinterval}.

\begin{lemma}\label{lemma:zeroerror}
Let $f(x^m)$ be a symmetric function of binary variables and
$\{[a_v,b_v]\}_{v=1}^{v_{\max}}$ the collection of all the maximal
$f'$-monochromatic intervals associated with $f$. Let $X^m$ be $m$
independent nontrivial Bernoulli sources. For any $r$-round,
block-length $n$ code\footnote{The results of
Lemma~\ref{lemma:zeroerror} hold for not only the multiround block
coding strategy described in Definition~\ref{def:code} but also for
the class of collision-free coding strategies defined in
\cite{Kumar2005}.}
%
for computing $f$ in a collocated network, if $\pP(\widehat{\mathbf
Z} \neq \mathbf Z)=0$, then given all the messages $M^t$, \emph{for
every sample} $i \in [1,n]$, the following conditions hold. (i)
There exists $v(M^t,i)\in [1,v_{\max}]$ such that
$S(i):=\left(\sum_{j=1}^m X_j(i)\right)\in [a_{v(M^t,i)},
b_{v(M^t,i)}])$. (ii) There exist $\mathcal
K_1(M^t,i)\subseteq[1,m]$ and $\mathcal K_0(M^t,i)\subseteq[1,m]$
such that: $\mathcal K_1(M^t,i) \bigcap \mathcal K_0(M^t,i)=\{ \},
|\mathcal K_1(M^t,i)|\geq a_{v(M^t,i)}, |\mathcal K_0(M^t,i)|\geq
m-b_{v(M^t,i)}$, and $\forall j\in \mathcal K_1(M^t,i), \forall j'
\in \mathcal K_0(M^t,i), X_j(i)=1, X_{j'}(i)=0$.
\end{lemma}

The proof of Lemma~\ref{lemma:zeroerror}, given in
Appendix~\ref{app:proofzeroerror}, is similar in structure to those
of Lemmas~\ref{lem:rectangle} and \ref{lemma:individualinterval}.



\emph{Example: (Parity function)} Let $f(x^m)=\left( \bigoplus_{i=1}^m
x_i\right)$ be the Boolean XOR function (parity) of $m$ binary
variables. Then $f'^{-1}(0)=\bigcup_{i=0}^{\lfloor m/2\rfloor} \{ 2 i
\}$ and $f'^{-1}(1)=\bigcup_{i=1}^{\lceil m/2\rceil} \{ 2 i-1
\}$. Thus for all $v \in [1,(m+1)]$, $a_v=b_v=(v-1)$, and all the
$f'$-monochromatic intervals are singletons. For every sample $i \in
[1,n]$, if $f$ is computed with zero error,
Lemma~\ref{lemma:zeroerror}(i) shows that the sink ends up knowing
$S(i)$ exactly, because every interval is now a singleton. In
addition, Lemma~\ref{lemma:zeroerror}(ii) shows that the sink will
also identify $S(i)$ source nodes which observe ones and $(m-S(i))$
source nodes which observe zeros. Therefore the sink essentially needs
all the raw data from all the source nodes in order to compute the
parity function in a collocated network.

\subsection{Bounds for minimum sum-rate}\label{subsec:bounds}

Returning to the block-error probability criterion,
Lemma~\ref{lemma:individualinterval} leads to the following bounds for
$R_{sum,r}$ when $X^m \sim$ iid Bernoulli$(p)$, that is, $\forall
i\in[1,m]$, $\pi_{i} = p$.

\begin{theorem}\label{thm:lowersymmfuncBerp}
Let $f(x^m)$ be a symmetric function of binary variables and
$\{[a_v,b_v]\}_{v=1}^{v_{\max}}$ the collection of all the maximal
$f'$-monochromatic intervals associated with $f$.  If $X^m \sim$ iid
Bernoulli$(p)$, $p\in (0,1)$, then for all $r\in \zZ^+$, (i)
\[R_{sum,r}\geq m h(p)- \! \sum_{v=1, a_v\neq b_v}^{v_{\max}} \! (b_v-a_v) h\left({\textstyle \frac{\eE(S|S\in
[a_v,b_v])-a_v}{b_v-a_v} }\right) \pP(S\in [a_v,b_v]),\] (ii) \quad
$\displaystyle R_{sum,r} \geq m h(p)
\max_{v \in [1,v_{\max}]} [ \pP(S\leq b_v) \pP(S> b_v)],$\\
(iii) \quad $\displaystyle R_{sum,r}\leq h(p)
\sum_{v=1}^{v_{\max}}\left(\frac{a_v}{p}+\frac{m-b_v}{1-p}\right)
\pP(S\in [a_v,b_v]),$\\
where $h(\cdot)$ is the binary entropy
function.
\end{theorem}

\emph{Remark:} The minimum sum-rate for ``data downloading'' where
all source samples are to be reproduced at the sink is $H(X^m)=m
h(p)$. Theorem~\ref{thm:lowersymmfuncBerp}(ii) explicitly bounds the
efficiency of multiround broadcasting relative to data downloading.
Since (ii) is proved by relaxing the lower bound in (i), the right
side of (ii) is not greater than that of (i).


\subsection{Scaling law of minimum sum-rate}\label{subsec:scaling}

Consider a sequence of problems, where in the $m$-th problem, $m\in
\zZ^+$, $m$ source nodes observe Bernoulli$(p_m)$ source samples
$\mathbf X^m$ which are iid both across samples and across nodes
and $f_m$ is the desired function. Let $R_{sum,r,m}$ be the minimum
sum-rate of the $m$-th problem. The scaling law of $R_{sum,r,m}$
with respect to $m$ is explored in the following cases.

\emph{Case 1:} We need to use the following fact.
\begin{fact}\label{fact:case1}
For any $\epsilon \in (0,1/2)$, if $\max_{v \in [1,v_{\max}]} [
\pP(S\leq b_v) \pP(S> b_v)]<\epsilon (1-\epsilon)$, then
$\exists~v^*\in [1,v_{\max}]$ such that
$\pP(S\in[a_{v^*},b_{v^*}])>1-2\epsilon$.
\end{fact}
\begin{IEEEproof}
For any $\epsilon \in (0,1/2)$, if $\max_{v \in [1,v_{\max}]} [
\pP(S\leq b_v) \pP(S> b_v)]<\epsilon (1-\epsilon)$, then $\forall v,
\pP(S\leq b_v)\in [0,\epsilon)\bigcup(1-\epsilon,1]$, which in turn
implies that for $v^*=\min \{v|\pP(S\leq b_v)>1-\epsilon\}$,
$\pP(S\in[a_{v^*},b_{v^*}])=\pP(S\leq b_{v^*})-\pP(S\leq
b_{v^*-1})>1-2\epsilon$ holds.
\end{IEEEproof}

If $\exists\epsilon\in (0,1/2)$ such that for every maximal
$f'_m$-monochromatic interval $[a_{v,m},b_{v,m}]$, $\pP(S\in
[a_{v,m},b_{v,m}])\leq (1- 2 \epsilon)$, then due to
Fact~\ref{fact:case1}, $\max_{v \in [1,v_{\max}]} [ \pP(S\leq b_v)
\pP(S> b_v)]\geq \epsilon (1-\epsilon)$. Then due to
Theorem~\ref{thm:lowersymmfuncBerp}(ii), $R_{sum,r,m}\geq
\epsilon(1-\epsilon) m h(p_m)$, which implies that
$R_{sum,r,m}=\Theta(m h(p_m))$ and data downloading is orderwise
optimal. Conversely, if $R_{sum,r,m}=o(m h(p_m))$, then due to
Theorem~\ref{thm:lowersymmfuncBerp}(ii), $\max_{v \in [1,v_{\max}]}
[ \pP(S\leq b_v) \pP(S> b_v)]\rightarrow 0$ as $m\rightarrow
\infty$. Therefore there exists a vanishing sequence $\{ \epsilon_m
\}_{m \in \zZ^+}$ such that $\max_{v \in [1,v_{\max}]} [ \pP(S\leq
b_v) \pP(S> b_v)]<\epsilon_m (1-\epsilon_m)$ holds for the $m$-th
problem. Due to Fact~\ref{fact:case1}, there exists a sequence of
maximal $f'_m$-monochromatic intervals $\{
[a_{v^*,m},b_{v^*,m}]\}_{m\in \zZ^+}$ such that $\pP(S \in
[a_{v^*,m},b_{v^*,m}])\rightarrow 1$ as $m\rightarrow \infty$. In
other words, multiround computation of symmetric functions of iid
binary sources in collocated networks is orderwise more efficient
than data downloading only if each sample of $f_m$ is determined
with a probability which tends to one as $m\rightarrow
\infty$.\footnote{We cannot, however, let nodes send nothing and set
the output of the sink to be the determined function value because
then, for each $m$ the probability of \emph{block error} will tend
to one with increasing block-length violating
Definition~\ref{def:rateregion}.}

\emph{Case 2: ($p_m=1/2$)} For any symmetric function of iid
Bernoulli$(1/2)$ sources, let
$\rho:=m-\sum_{v=1}^{v_{\max}}\left(b_v-a_v\right) \pP(S\in
[a_v,b_v])$. Theorem~\ref{thm:lowersymmfuncBerp}(i) and (iii) imply
that $\rho \leq R_{sum,r}\leq R_{sum,1}\leq 2 \rho$. This shows that
multiround computation can at most halve the minimum sum-rate of
one-round computation. Since $\rho$ can be easily computed using the
binomial distribution,
$R_{sum,r}$ can be easily evaluated within a factor of $2$ for all
$r\in \zZ^+$.

\emph{Case 3: (Type-sensitive functions, $p_m=1/2$)} A sequence of
symmetric functions $\{f_m\}_{m\in \zZ^+}$ of binary variables is
type-sensitive if $\exists~\gamma\in(0,1)$ and $\bar m\in \zZ^+$ such
that $\forall m>\bar m$, for every $f'_m$-monochromatic interval
$[a_{v,m},b_{v,m}]$, $(b_{v,m}-a_{v,m})< \gamma \bar m$ (defined in
\cite{Kumar2005}, adapted to our notation). For example, the sum,
mode, and parity functions are type-sensitive. For iid
Bernoulli$(1/2)$ sources, it can be shown that $R_{sum,r,m}=\Theta(m)$
by applying Theorem~\ref{thm:lowersymmfuncBerp}(i). {\em Remark:} For
the zero-error criterion, the minimum worst-case sum-rate is also
$\Theta(m)$\cite{Kumar2005}.

\emph{Case 4: (Type-threshold functions)} A sequence of symmetric
functions $\{f_m\}_{m\in \zZ^+}$ of binary variables is
type-threshold, if there exist $\theta_0, \theta_1\in \nN$ such that
$[\theta_1, m-\theta_0]$ is $f'_m$-monochromatic for every $m\geq
\theta_0+\theta_1$ (defined in \cite{Kumar2005}, adapted to our
notation). For example, the minimum and maximum functions are
type-threshold. (i) If $p_m=p$, then $\pP(S\in [\theta_1,
m-\theta_0])\rightarrow 1$ exponentially fast as $m\rightarrow
\infty$. By applying Theorem~\ref{thm:lowersymmfuncBerp}(i) and
(iii), we have $R_{sum,r,m}=\Theta(1)$, which is orderwise less than
$H(m h(p_m))=\Theta(m)$. (ii) If $p_m=1/m$ and
$f_m(x^m)=\max_{i=1}^m x_i$, then $a_1=b_1=0$, $a_2=1$, $b_2=m$, and
$\lim_{m\rightarrow \infty} \pP(S\leq 0)\pP(S>0)=e^{-1}(1-e^{-1})$,
due to Theorem~\ref{thm:lowersymmfuncBerp}(ii),
$R_{sum,r,m}=\Theta(m h(p_m))=\Theta(\log m)$. {\em Remark:} For the
zero-error criterion, the minimum worst-case sum-rate is
$\Theta(\log m)$\cite{Kumar2005}.

\subsection{Comparison to cut-set bounds}\label{subsec:cutset}

How do the bounds given in Sec.~\ref{subsec:bounds} behave in
comparison to bounds based on cut-sets? We will show that in some
cases they are orderwise tighter than cut-set bounds and in some cases
they coincide with them.

For any subset $\mathcal S\subseteq [1,m]$, let $\mathcal
S^c:=[1,m]\setminus \mathcal S$. We can formulate a two-terminal
interactive function computation problem with alternating message
transfers \cite{ISIT08} by regarding the set of source nodes in
$\mathcal S$ as supernode-$\mathcal S$ and the other source nodes and
the sink as supernode-$\mathcal S^c$. The sources $\{X_i\}_{i\in
\mathcal S}$ and $\{X_i\}_{i\in \mathcal S^c}$ are available to
supernode-$\mathcal S$ and supernode-$\mathcal S^c$ respectively and
the function $f(X^m)$ is to be computed at supernode-$\mathcal
S^c$. Let $\mathcal R_{\mathcal S, \mathcal S^c}$ denote the directed
sum-rate {\em region} of the two-terminal problem, which is the set of
tuples $(R_{\mathcal S\rightarrow \mathcal S^c}, R_{\mathcal S^c
\rightarrow \mathcal S})$ such that $R_{\mathcal S\rightarrow \mathcal
S^c}$ and $R_{\mathcal S^c\rightarrow \mathcal S}$ are admissible
directed sum-rates from $\mathcal S$ to $\mathcal S^c$ and from
$\mathcal S^c$ to $\mathcal S$ respectively, for two-terminal
interactive function computation with $t'$ alternating messages where
$t' \leq 2mr$ is the minimum number of messages needed in the
two-terminal problem to simulate the multiround code.

For any multiround code for a collocated network, for every $i \in
[1,m]$, let $r_i$ denote the sum-rate of the messages broadcasted by
node-$i$. This code can be mapped into a two-terminal interaction
code for the two-terminal problem described above, which generates
the same computation result. The directed sum-rate tuples is
$(R_{\mathcal S\rightarrow \mathcal S^c},R_{\mathcal S^c\rightarrow
\mathcal S}) = \left(\sum_{i\in \mathcal{S}} r_i, \sum_{i\in
\mathcal{S}^c} r_i\right)$, which should belong to the directed
sum-rate region of the two-terminal problem. This leads to the
following cut-set bound.

\begin{theorem}\label{thm:cutset}\emph{(cut-set bound)} For all $r\in
\zZ^+$, \begin{equation} R_{sum,r}\geq R_{cut}:=\min_{\scriptstyle
\forall \mathcal{S}\subseteq [1,m],\ \left(\sum_{i\in \mathcal{S}}
r_i, \sum_{i\in \mathcal{S}^c} r_i\right)\in \mathcal R_{\mathcal S,
\mathcal S^c} \atop \scriptstyle \forall i\in[1,m],\ r_i\geq 0}
\sum_{i=1}^m r_i.\label{eqn:cutset}\end{equation}
\end{theorem}

One could also consider a different type of cut-set bound:
\begin{equation*} R_{sum,r}\geq R'_{cut}:= \max_{\mathcal{S}\subseteq [1,m]} R_{\mathcal S,
\mathcal S^c}^{sum}:= \max_{\mathcal{S}\subseteq [1,m]}
\left(\min_{\scriptstyle \left(\sum_{i\in \mathcal{S}} r_i,
\sum_{i\in \mathcal{S}^c} r_i\right) \in \mathcal R_{\mathcal S,
\mathcal S^c} \atop \scriptstyle \forall i\in[1,m],\ r_i\geq 0}
\sum_{i=1}^m r_i \right),\end{equation*} where $R_{\mathcal S,
\mathcal S^c}^{sum}$ is called the bi-directional minimum sum-rate
of the two-terminal problem given by the cut-set $\mathcal S$. Note
that $R'_{cut}\leq R_{cut}$. In fact, $R'_{cut}$ can be orderwise
looser than $R_{cut}$. For example, for the problem in
Prop.~\ref{prop:cutsettight}, $R_{cut}\geq m$ and $R'_{cut}=1$.

\begin{proposition}\label{prop:cutsettight}
If $X^m\sim $ iid Bernoulli$(1/2)$ and
$f_m(x^m)=\left(\bigoplus_{i=1}^m x_i\right)$, then $R_{cut}\geq m$.
\end{proposition}
\begin{IEEEproof}
For any $i\in[1,m]$, if $\mathcal S=\{i\}$, by applying the cut-set
bound for the two-terminal interactive function computation problem
\cite[Corollary~1(ii)]{ISIT08}, we have $\forall \left(r_i,
\sum_{j\neq i} r_j\right)\in \mathcal R_{\mathcal S,\mathcal S^c}$,
$r_i \geq H(f_m(X^m)|\{X_k\}_{k\in \mathcal S^c})=H(X_i)=1$. Adding
the $m$ inequalities $r_i \geq 1$ for all $i \in [1,m]$, we have
$R_{cut}\geq m$.
\end{IEEEproof}

Since $m$ is an admissible sum-rate for the problem stated in
Prop.~\ref{prop:cutsettight}, the cut-set bound is tight. Note that
Theorem~\ref{thm:lowersymmfuncBerp}(i) also gives the same bound
$R_{sum,r}\geq m$. However, in the following case, the cut-set bound
is orderwise loose.

\begin{proposition}\label{prop:cutsetloose}
If $X^m\sim $ iid Bernoulli$(1/2)$ and $f_m(x^m)=\min_{i=1}^m x_i$,
then $R_{cut}\leq 3m/(2^{m/2})$.
\end{proposition}
\begin{IEEEproof} It is sufficient to show that $\forall i
\in [1,m]$, $r_i=3/(2^{m/2})$ is feasible for the minimization
problem in (\ref{eqn:cutset}), which requires showing $\forall
\mathcal S\subseteq [1,m]$, $\left(\sum_{i\in \mathcal{S}} r_i,
\sum_{i\in \mathcal{S}^c} r_i\right)=(3|\mathcal
S|/(2^{m/2}),3|\mathcal S^c|/(2^{m/2}))\in \mathcal R_{\mathcal S,
\mathcal S^c}$. Let $\mathbf Y_{\mathcal S}:=\left(\min_{i\in
\mathcal S} X_i(k) \right)_{k=1}^n \sim$ iid~$p_{Y_{\mathcal
S}}\sim$ Bernoulli$(1/(2^{|\mathcal S|}))$ and $\mathbf Y_{\mathcal
S^c}:=\left(\min_{i\in \mathcal S^c} X_i(k)\right)_{k=1}^n \sim$
iid~ $p_{Y_{S^c}} \sim$ Bernoulli$(1/(2^{|\mathcal S^c|}))$. The
computation of the two-terminal problem can be performed by the
following two schemes. (i) (One-message scheme) Supernode-$\mathcal
S$ sends $\mathbf Y_{\mathcal S}$ to supernode-$\mathcal S^c$ at the
rate $H(Y_{\mathcal S})$. Therefore $(H(Y_{\mathcal S}),0)\in
\mathcal R_{\mathcal S, \mathcal S^c}$, which implies that $\mathcal
R^1:=[H(Y_{\mathcal S}),\infty)\times [0,\infty) \subseteq \mathcal
R_{\mathcal S, \mathcal S^c}$. (ii) (Two-message scheme)
Supernode-$\mathcal S^c$ sends $\mathbf Y_{\mathcal S^c}$ to
supernode-$\mathcal S$ at the rate $H(Y_{\mathcal S^c})$. Then
supernode-$\mathcal S$ computes the samplewise minimum of $\mathbf
Y_{\mathcal S}$ and $\mathbf Y_{\mathcal S^c}$, and sends it back to
supernode-$\mathcal S^c$ with $\mathbf Y_{\mathcal S^c}$ as side
information available to both supernodes, at the rate
$H(\min(Y_{\mathcal S},Y_{\mathcal S^c})| Y_{\mathcal S^c})$.
Therefore $(H(\min(Y_{\mathcal S},Y_{\mathcal S^c})| Y_{\mathcal
S^c}),H(Y_{\mathcal S^c}))\in \mathcal R_{\mathcal S, \mathcal
S^c}$, which implies that $\mathcal R^2:=[H(\min(Y_{\mathcal
S},Y_{\mathcal S^c})| Y_{\mathcal S^c}),\infty)\times [H(Y_{\mathcal
S^c}),\infty) \subseteq \mathcal R_{\mathcal S, \mathcal S^c}$. By
evaluating the entropies, it can be shown that, if $|\mathcal S|\geq
m/2$, then $(\sum_{i\in \mathcal{S}} r_i, \sum_{i\in \mathcal{S}^c}
r_i)\in \mathcal R^1$, otherwise $(\sum_{i\in \mathcal{S}} r_i,
\sum_{i\in \mathcal{S}^c} r_i)\in \mathcal R^2$. Therefore
$(\sum_{i\in \mathcal{S}} r_i, \sum_{i\in \mathcal{S}^c} r_i)\in
\mathcal R_{\mathcal S, \mathcal S^c}$.

The detailed steps only for ourselves: (will be deleted in the final
draft) If $|\mathcal S|\geq m/2$, then
\[H(Y_{\mathcal S})=h\left(\frac{1}{2^{|\mathcal S|}}\right)\leq \frac{\log_2(e 2^{|\mathcal S|})}{2^{|\mathcal S|}}
\leq \frac{\log_2(8^{|\mathcal S|})}{2^{|\mathcal S|}}\leq
\frac{3|\mathcal S|}{2^{m/2}},
\]
where the first inequality is because $h(p)\leq p \log_2(e/p)$ and
the second inequality is because $e< 4^{|\mathcal S|}$. Therefore
$(\sum_{i\in \mathcal{S}} r_i, \sum_{i\in \mathcal{S}^c}
r_i)=(3|\mathcal S|/(2^{m/2}),3|\mathcal S^c|/(2^{m/2}))\in \mathcal
R^1$. If $|\mathcal S|< m/2$, then $H(Y_{\mathcal S})\leq 3|\mathcal
S^c|/(2^{m/2})$. If $1\leq |\mathcal S|< m/2$,
\[H(\min(Y_{\mathcal S},Y_{\mathcal
S^c})| Y_{\mathcal S^c})=\frac{1}{2^{|\mathcal S^c|}}
h\left(\frac{1}{2^{|\mathcal S|}}\right) \leq \frac{3|\mathcal
S|}{2^{m}} \leq \frac{3|\mathcal S|}{2^{m/2}}.
\]
Otherwise ($|\mathcal S|=0$), $H(\min(Y_{\mathcal S},Y_{\mathcal
S^c})| Y_{\mathcal S^c})=0=3|\mathcal S|/(2^{m/2})$. Therefore
$(\sum_{i\in \mathcal{S}} r_i, \sum_{i\in \mathcal{S}^c}
r_i)=(3|\mathcal S|/(2^{m/2}),3|\mathcal S^c|/(2^{m/2}))\in \mathcal
R^2$.
\end{IEEEproof}

Since the problem considered in Prop.~\ref{prop:cutsetloose} is a
special case of Case~4(i), due to Theorem~\ref{thm:lowersymmfuncBerp}
we have $R_{sum,r,m}=\Theta(1)$.  Therefore the exponentially
vanishing cut-set bound given by Theorem~\ref{thm:cutset} is orderwise
loose.

\section{Concluding Remarks}

We studied function computation in collocated networks using a
distributed block source coding framework. We showed that in computing
symmetric functions of binary sources, the sink will inevitably obtain
certain additional information which is not part of the problem
requirement. Leveraging this conceptual understanding we developed
bounds for the minimum sum-rate and showed that they can be better
than cut-set bounds by orders of magnitude. Directions for future work
include characterizing the scaling law of the minimum sum-rate for
large source alphabets and general multihop networks.

\appendices
\renewcommand{\theequation}{\thesection.\arabic{equation}}
\setcounter{equation}{0}

\section{\label{app:converse}Converse proof of Theorem~\ref{thm:rateregion}}
Suppose a rate tuple $(R_1, \ldots, R_t)$ is admissible for
$r$-round function computation. By Definition~\ref{def:rateregion},
$\forall \epsilon>0, \exists \bar n(\epsilon,t)$, such that $\forall
n>\bar n(\epsilon)$, there exists an $r$-round distributed source
code satisfying $\forall j\in[1,t], (1/n) \log_2 |\mathcal M_j|<
R_j+\epsilon$ and $\pP(\mathbf Z\neq \widehat{\mathbf Z})<
\epsilon$. Define auxiliary random variables as follows: $\forall
k\in[1,n], \ U_1(k) :=\{M_1,X^m(k-)\}$ \footnote{$A(k-)$ means
$\{A(1),\ldots,A(k-1)\}$, and $A(k+)$ means
$\{A(k+1),\ldots,A(m)\}$.}, and for $\forall i\in[2,t]$, $U_i :=
M_i$.

{\em Information inequalities:} For the first rate, we have
\begin{eqnarray}
n(R_1+\epsilon) &\geq& H(M_1) \nonumber\\
&\geq& I(\mathbf X_1;M_1|\mathbf X_2^m) \nonumber\\
&=& H(\mathbf X_1|\mathbf X_2^m)-H(\mathbf X_1|M_1,\mathbf X_2^m) \nonumber\\
&=&\sum_{k=1}^n (H(X_1(k)) - H(X_1(k)|X_1(k-),
M_1,\mathbf X_2^m)) \nonumber\\
&\geq &\sum_{k=1}^n (H(X_1(k)) - H(X_1(k)|M_1,X^m(k-)))\nonumber\\
&= &\sum_{k=1}^n I(X_1(k);U_1(k)).\label{eqn:rate1}
\end{eqnarray}

For the $i$-th rate, $i\in[2,t]$, let $j=(i \!\! \mod m)$.
\begin{eqnarray}
n(R_i+\epsilon) &\geq& H(M_i) \nonumber\\
&\geq& I(\mathbf X_j;M_i|M^{i-1},\mathbf X^{j-1},\mathbf X_{j+1}^m) \nonumber\\
&=& H(\mathbf X_j|M^{i-1},\mathbf X^{j-1},\mathbf X_{j+1}^m)-H(\mathbf X_j|M^{i},\mathbf X^{j-1},\mathbf X_{j+1}^m) \nonumber\\
&=&\sum_{k=1}^m (H(X_j(k)|X_j(k-),M^{i-1},\mathbf X^{j-1},\mathbf
X_{j+1}^m) \nonumber\\ && \ \ \ \ - H(X_j(k)|X_j(k-),M^i,\mathbf
X^{j-1},\mathbf
X_{j+1}^m)) \nonumber\\
&\stackrel{(a)}{=}&\sum_{k=1}^n (H(X_j(k)|X^m(k-),M^{i-1}) \nonumber
\\ && \ \ \ \ - H(X_j(k)|X_j(k-),M^i,\mathbf X^{j-1},\mathbf
X_{j+1}^m))\nonumber\\
&\geq & \sum_{k=1}^n (H(X_j(k)|X^m(k-),M^{i-1}) \nonumber \\ && \ \
\ \ -H(X_j(k)|X^m(k-),M^{i}))\nonumber\\
&= &\sum_{k=1}^n I(X_j(k);U_i|U_1(k),U_2^{i-1}).\label{eqn:ratei}
\end{eqnarray}
Step (a) is because the Markov chain $X_j(k)- (M^{i-1}, X^m(k-))-
(X^{j-1}(k),X^{j-1}(k+),X_{j+1}^m(k),X_{j+1}^m(k+))$ holds for each
$i\in[1,t]$ and $k\in[1,n]$.

Due to the condition $\pP(\mathbf Z \neq \mathbf{\widehat Z}) \leq
\epsilon$ and the Fano's inequality\cite{CoverThomas}, we have
\begin{eqnarray}
\lefteqn{ h_2(\epsilon)+ \epsilon \log (|\mathcal Z|^n-1)}\nonumber
 \\&\geq& H(\mathbf{Z}| M^t)  \nonumber\\
&=& \sum_{k=1}^n H(Z(k)| Z(k-), M^t)  \nonumber\\
&\geq & \sum_{k=1}^n H(Z(k)| Z(k-), M^t, X^m(k-))  \nonumber\\
&\stackrel{(b)}{=}& \sum_{k=1}^n H(Z(k)| M^t, X^m(k-))  \nonumber\\
&=&\sum_{k=1}^n H(f(X^m(k))|U_1(k),U_2^t).\label{eqn:entropy}
\end{eqnarray}
Step (b) is because $Z(k-)$ is a function of $X^m(k-)$.

{\em Timesharing:} We introduce a timesharing random variable $Q$
taking values in $[1,n]$ equally likely, which is independent of all
the other random variables. For each $i\in[1,m]$, define
$X_i:=X_i(Q)$, and $U_1:=(U_1(Q),Q)$. (\ref{eqn:rate1}) becomes
\begin{eqnarray}
R_1+\epsilon & \geq & \frac{1}{n} \sum_{k=1}^n I(X_1(k);U_1(k)) \nonumber\\
&=&I(X_1(Q);U_1(Q)|Q)\nonumber\\ &\stackrel{(c)}{=}& I(X_1(Q);U_1(Q),Q)\nonumber\\
&=& I(X_1;U_1), \label{eqn:rate1timesharing}
\end{eqnarray}
where step (c) is because $I(X_1(Q);Q)=0$, which is in turn implied
by: (1), $Q$ is independent of all the other random variables, and
(2), the distribution of $X_1(Q)\sim p_{X_1}$ does not depend on
$Q$. Similarly, (\ref{eqn:ratei}) and (\ref{eqn:entropy}) become
\begin{eqnarray}
\forall i\in[2,t],\ \  R_i+\epsilon &\geq&  I(X_j;U_i|U^{i-1}),
\label{eqn:rateitimesharing}\\
\frac{1}{n}h_2(\epsilon)+ \epsilon \log |\mathcal {Z}| &\geq&
H(f(X^m)|U^t),\label{eqn:entropytimesharing}
\end{eqnarray}

Concerning the Markov chains, one can verify that
$U_1(k)-X_1(k)-X_2^m(k)$ and
$U_i-(U_1(k),U_2^{i-1},X_j(k))-(X^{j-1}(k),X_{j+1}^m(k))$ form
Markov chains for each $i\in[2,t]$, $k\in[1,n]$ and $j=(i \!\! \mod
m)$, which imply that $U_i-(U^{i-1},X_j)-(X^{j-1},X_{j+1}^m)$ forms
a Markov chain for each $i\in [1,t]$ and $j=(i \!\! \mod m)$.

{\em Cardinality bounds:} The cardinalities of $\mathcal U_i$ can be
bounded as for the rate region of the two-terminal interaction
problem\cite{ISIT08}. But they are omitted here.

{\em Taking limits:} As in \cite{ISIT08}, we consider a sequence
$\{\epsilon_l\}$ which goes to zero as $l$ tends to infinity. Due to
the continuity of conditional mutual information and conditional
entropy measures, all the $\epsilon$'s in
(\ref{eqn:rate1timesharing})-(\ref{eqn:entropytimesharing}) vanish
and thus $(R_1,\ldots,R_t) \in \mathcal R_r$.


\section{Proof of Lemma~\ref{lemma:zeroerror}}
\label{app:proofzeroerror}

For any $r$-round block-length $n$ code $(e^t,g)$ for computing $f$
without any error in a collocated network, for every realization
$m^t$ of messages $M^t$ and for every sample $i\in [1,n]$, let
$\mathcal A(m^t,i):=\{x^m (i) | p_{X^m(i),M^t}(x^m(i),m^t)>0 \}$,
i.e., the set of all possible $i$-th source samples that are
consistent with messages $m^t$. We first show that $\mathcal
A(m^t,i)$ is an $f$-monochromatic rectangle in
$\bigotimes_{i=1}^{m}\mathcal X_i$, which is similar to the
statement of Lemma~\ref{lem:rectangle}. Due to
Definition~\ref{def:code}, $\mathcal A(m^t,i)=\{x^m (i) |
\exists~x^m(1),\ldots,x^m(i-1),x^m(i-1),\ldots,x^m(n)$ such that
$\forall j\in [1,t], k=(j \!\mod m), e_j(\mathbf
x_k,m^{j-1})=m_j\}$, where $\mathbf x_k$ stands for
$(x_k(1),\ldots,x_k(n))$. For every $k\in [1,m]$, let $\mathcal
S_k(m^t,i):=\{x_k (i) |
\exists~x_k(1),\ldots,x_k(i-1),x_k(i-1),\ldots,x_k(n)$ such that
$\forall \rho \in [0,r-1], j=k+\rho m, e_j(\mathbf
x_k,m^{j-1})=m_j\}$. Since $\mathcal S_k(m^t,i)$ contains all the
constraints in $\mathcal A(m^t,i)$ related to source node $k$, we
have $\mathcal A(m^t,i)=\bigotimes_{k=1}^{m} \mathcal S_k(m^t,i)$.
Therefore $\mathcal A(m^t,i)$ is a rectangle in
$\bigotimes_{i=1}^{m}\mathcal X_i$. Since the code computes $f$
without any error for any inputs, $\mathcal A(m^t,i)$ is
$f$-monochromatic.

The rest steps are parallel to the proof of
Lemma~\ref{lemma:individualinterval}. For all possible messages
$m^t$, $\mathcal A(m^t,i)$ is nonempty and $\mathcal S_i(m^t,i)$ is
either $\{0\}$ or $\{1\}$ or $\{0,1\}$. Let $\mathcal
K_1(m^t,i):=\{i~|~ \mathcal S_i(m^t,i)=\{1\} \}$ and $\mathcal
K_0(m^t,i):=\{i~|~ \mathcal S_i(m^t,i)=\{0\} \}$. Let
$\alpha(m^t,i):=|\mathcal K_1(m^t,i)|$ and $\beta(m^t,i):= m -
|\mathcal K_0(m^t,i)|$. It can be shown that the projection of
$\mathcal{A}(m^t,i)$ under the linear transformation
$s=\left(\sum_{i=1}^m x_i\right)$ given by
$\mathcal{A}'(m^t,i):=\left\{\left(\sum_{i=1}^m
x_i(i)\right)|p_{X^m(i),M^t}(x^m(i),m^t)>0 \right\}$ is an
$f'$-monochromatic interval $[\alpha(m^t,i),\beta(m^t,i)]$. Since
$\{[a_v,b_v]\}_{v=1}^{v_{\max}}$ is the collection of all the
maximal $f'$-monochromatic intervals, $\exists~v(m^t,i)\in
[1,v_{\max}]$ such that $\mathcal{A}'(m^t,i)\subseteq
[a_{v(m^t,i)},b_{v(m^t,i)}]$. Therefore (i) holds. Since
$\mathcal{A}'(m^t,i)=[\alpha(m^t,i),\beta(m^t,i)] \subseteq
[a_{v(m^t,i)},b_{v(m^t,i)}]$, we have $\alpha(m^t,i)\geq
a_{v(m^t,i)}$ and $\beta(m^t,i)\leq b_{v(m^t,i)}$. Therefore (ii)
holds.

\section{Proof of Theorem~\ref{thm:lowersymmfuncBerp}}
\label{app:prooflowersymmfuncBerp}

\noindent {\em Proof of Theorem~\ref{thm:lowersymmfuncBerp} (i):} We
first state a lemma.
\begin{lemma}\label{lemma:fact} If $Y^k \in \{0,1\}^k$ is a random vector and
$S_Y=\sum_{i=1}^k Y_i$, then $H(Y^k)\leq k
h\left(\eE(S_Y)/k\right)$.
\end{lemma}
\begin{IEEEproof}
\[H(Y^k)\leq \sum_{i=1}^k H(Y_i)=\sum_{i=1}^k h(\eE(Y_i))
\stackrel{(a)}{\leq} k h\left({\textstyle\frac{\sum_{i=1}^k
\eE(Y_i)}{k}}\right) =k h\left({\textstyle
\frac{\eE(S_Y)}{k}}\right).\] Step (a) is due to the concavity of
$h(x)$ and the Jensen's inequality.
\end{IEEEproof}

Define an auxiliary random variable $V$ by $V:=v$ if and only if
$S\in [a_v,b_v]$. Lemma~\ref{lemma:individualinterval}(i) implies
that $p_{V|U^t}(v|u^t)=\delta_{v,v(u^t)}$ and thus $V=v(U^t)$.
 Due to Corollary~\ref{cor:sumrate}, we have
\begin{eqnarray}
R_{sum,r} &=& \min_{U^t} I(X^m;U^t)\nonumber\\
&=& \min_{U^t} [H(X^m)-H(X^m|U^t,V)]\nonumber\\
&=& m h(p)-\max_{U^t} \sum_{v=1}^{v_{\max}}
H(X^m|U^t,V=v)P_{V}(v),\label{eqn:eqn0AppLB}
\end{eqnarray}
where $U^t$ are subject to all the Markov chain and conditional
entropy constraints in (\ref{eqn:rateregion}). Due to
Lemma~\ref{lemma:individualinterval}(ii), given that $U^t = u^t\in
\supp(p_{U^t})$, $\forall i\in \mathcal{K}_1(u^t)$, $X_i=1$, and
$\forall i'\in \mathcal{K}_0(u^t)$, $X_{i'}=0$. Therefore there are
at most
$(m-|\mathcal{K}_1(u^t)+\mathcal{K}_0(u^t)|)=(b_{v(u^t)}-a_{v(u^t)})$
undetermined sources. When $a_{v(u^t)}=b_{v(u^t)}$, all sources are
determined. Therefore, for those $v$ satisfying $a_v=b_v$,
$H(X^m|U^t,V=v)=0$ holds. For other $v$'s we have
\begin{eqnarray}
\lefteqn{H(X^m|U^t,V=v)}\nonumber\\
&=& \sum_{u^t}
H(X^m|U^t=u^t,V=v)p_{U^t|V}(u^t|v)\nonumber\\
&\stackrel{(b)}{\leq}& (b_v-a_v) \sum_{u^t}
h\left(\eE\left(\frac{S-a_v}{b_v-a_v}\Bigg|U^t=u^t,V=v  \right)\right)p_{U^t|V}(u^t|v)\nonumber\\
&\stackrel{(c)}{\leq}& (b_v-a_v)
h\left(\eE\left(\frac{S-a_v}{b_v-a_v}\Bigg|V=v
\right)\right),\label{eqn:eqn1AppLB}
\end{eqnarray}
where the summations are through all $u^t\in
\supp(p_{U^t|V}(\cdot|v))$. Step (b) is due to
Lemma~\ref{lemma:fact}. Step (c) is due to the concavity of $h(x)$
and the Jensen's inequality. Combining (\ref{eqn:eqn0AppLB}) and
(\ref{eqn:eqn1AppLB}) leads to the statement of
Theorem~\ref{thm:lowersymmfuncBerp}(i).

\noindent {\em Proof of Theorem~\ref{thm:lowersymmfuncBerp} (ii):}
\begin{lemma}\label{lemma:combine} Let $S$ be a random variable taking integer
values. Let integers $a_1,b_1, a_2, b_2$ satisfy $a_1\leq b_1,
a_2\leq b_2$ and $a_2=b_1+1$. Then
\begin{eqnarray}\lefteqn{ \sum_{v=1,
a_v\neq b_v}^{2} \! (b_v-a_v) h\left({\textstyle \frac{\eE(S|S\in
[a_v,b_v])-a_v}{b_v-a_v} }\right) \pP(S\in [a_v,b_v])}\nonumber
\\&&\leq (b_2-a_1) h\left({\textstyle \frac{\eE(S|S\in
[a_1,b_2])-a_1}{b_2-a_1} }\right) \pP(S\in
[a_1,b_2]).\label{eqn:combine}\end{eqnarray}
\end{lemma}
\begin{IEEEproof}
If $a_1\neq b_1$ and $a_2\neq b_2$, it is sufficient to show that
\begin{eqnarray*}\lefteqn{
{\textstyle \frac{b_2-b_1}{b_2-a_1} h(0)\frac{ \pP(S\in
[a_1,b_1])}{\pP(S\in [a_1,b_2])}+\frac{a_2-a_1}{b_2-a_1} h(1)\frac{
\pP(S\in
[a_2,b_2])}{\pP(S\in [a_1,b_2])}}}\\
 \lefteqn{+\sum_{v=1, a_v\neq b_v}^{2} \!
{\textstyle \frac{b_v-a_v}{b_2-a_1} h\left({\frac{\eE(S|S\in
[a_v,b_v])-a_v}{b_v-a_v} }\right)\frac{ \pP(S\in
[a_v,b_v])}{\pP(S\in [a_1,b_2])}}} \\&&\leq (b_2-a_1)
h\left({\textstyle \frac{\eE(S|S\in [a_1,b_2])-a_1}{b_2-a_1}
}\right) \pP(S\in [a_1,b_2]),\end{eqnarray*} which is guaranteed by
the concavity of $h(x)$ and the Jensen's inequality. If $a_1=b_1$ or
$a_2=b_2$, drop the corresponding term in the summation, and the
proof continues to hold.
\end{IEEEproof}


The left hand side of (\ref{eqn:combine}) contains the terms in the
lower bound in Theorem~\ref{thm:lowersymmfuncBerp}(i) that
correspond to intervals $[a_1,b_1]$ and $[a_2,b_2]$. The right hand
side of (\ref{eqn:combine}) is a term corresponding to $[a_1,b_2]$.
Lemma~\ref{lemma:combine} shows that the lower bound given in
Theorem~\ref{thm:lowersymmfuncBerp}(i) will decrease when we combine
two adjacent intervals.

In order to prove Theorem~\ref{thm:lowersymmfuncBerp} (ii), it is
sufficient to show that $R_{sum,r}\geq m h(p) \pP(S\leq b_v)
\pP(S>b_v)$ holds for every $v \in [1,(v_{\max}-1)]$. For any
$v\in[1,v_{\max}-1]$, let $b:=b_v$, $p_0:=\pP(S\leq b)$ and
$p_1:=1-p_0$. If $b\neq 0$ and $b\neq m-1$, apply
Lemma~\ref{lemma:combine} to Theorem~\ref{thm:lowersymmfuncBerp}(i)
by combining all the intervals greater than $b$ into $[b+1,m]$ and
all the intervals not greater than $b$ into $[0,b]$, we have
\begin{equation}\label{eqn:rsumorder1}
R_{sum,r}\geq m h(p) - b p_0 h\left({\textstyle \frac{\eE(S|S\leq
b)}{b}}\right) -(m-b-1) p_1 h\left({\textstyle \frac{\eE(S|S\geq
b+1)-b-1}{m-b-1}}\right).\end{equation} If $b= 0$ or $b= m-1$, drop
the second or the third term on the right side of
(\ref{eqn:rsumorder1}) and the corresponding terms in the following
steps, and the proof continue to work. In order to show that the
right side of (\ref{eqn:rsumorder1}) is not less than $m h(p) p_0
p_1$, it is sufficient to show that
\begin{equation}\label{eqn:rsumorder2}\lambda_1 h(\alpha_1)+\lambda_2 h(\alpha_2)+\lambda_3 h(0)+\lambda_4 h(1)\leq h(p),\end{equation}
where
\[\lambda_1=\frac{b p_0}{m (1-p_0 p_1)},\ \ \ \  \lambda_2=\frac{(m-b-1)p_1}{m (1-p_0 p_1)},\]
\[\alpha_1=\frac{\eE(S|S\leq b)}{b},\ \ \ \  \alpha_2=\frac{\eE(S|S\geq b+1)-b-1}{m-b-1},\]
and $\lambda_3, \lambda_4$ are arbitrary real numbers.
(\ref{eqn:rsumorder2}) is guaranteed by the Jensen's inequality if
$\lambda_1 \alpha_1+\lambda_2 \alpha_2+\lambda_4=p$, $\sum_{i=1}^4
\lambda_i=1$ and $\lambda_i\geq 0$, $i=1,\ldots,4$. The first two
conditions imply that
\[
\lambda_3=\frac{p_0 (1-p)}{1-p_0 p_1}\left( \frac{m-b}{m-m p}- p_1
\right),\
 \lambda_4=\frac{p p_1}{1-p_0 p_1}\left( \frac{b+1}{m p}-
p_0\right).
\]

We need to verify that $\lambda_3\geq 0$ and $\lambda_4 \geq 0$. In
order to get $\lambda_4 \geq 0$, it is sufficient to verify that
\begin{equation}p_0=\pP(S\leq b)\leq \frac{b+1}{mp}.\label{eqn:lambda4nonneg}\end{equation}
When $b\geq mp-1$, (\ref{eqn:lambda4nonneg}) holds immediately. When
$b<m p-1$, we verify (\ref{eqn:lambda4nonneg}) as follows.

The probability mass function $p_S(s)$ is nondecreasing when $s \leq
\mu_S$, where $\mu_S=\lfloor m p+p\rfloor $ is the mode of the
binomial random variable $S$. Therefore $\{\pP(S\leq
b)\}_{b=0}^{\lfloor m p\rfloor}$ is a convex sequence \cite{Convex},
which implies that for all integers $0\leq b\leq \lfloor m
p\rfloor$, the point $(b,\pP(S\leq b))$ is below or on the line
segment joining the point $(0,\pP(S\leq 0))=(0,(1-p)^m)$ and the
point $(\lfloor mp \rfloor, \pP(S\leq \lfloor mp\rfloor) )$. Since
\[mp (1-p)^m \leq mp (1-p)^{m-1}\leq \sum_{k=0}^m {m \choose k} p^k
(1-p)^{m-1}=1,
\]
we have $(1-p)^m \leq 1/mp$. Also, $\pP(S\leq \lfloor mp\rfloor)\leq
1 <(\lfloor mp\rfloor+1)/mp$. Therefore the line segment joining
$(0,(1-p)^m)$ and $(\lfloor mp \rfloor, \pP(S\leq \lfloor mp\rfloor)
)$ is below the line segment joining $(0,1/mp)$ and $(\lfloor mp
\rfloor, (\lfloor mp\rfloor+1)/mp)$, which is the graph of the
function $(b+1)/mp$ when $0\leq b\leq \lfloor mp \rfloor$. Therefore
we have shown that (\ref{eqn:lambda4nonneg})  holds for $b\leq
\lfloor m p\rfloor$ and completed the proof for $\lambda_4 \geq 0$.
Similarly we have $\lambda_3 \geq 0$.

\noindent {\em Proof of Theorem~\ref{thm:lowersymmfuncBerp} (iii):}
Let $r=1$ and for each $i \in [1,m]$, define
\begin{eqnarray*}
U_i:= \left\{
\begin{array}{cl}
0, & \mbox{if } \exists\ v, s.t.\  N_1(U^{i-1})\geq a_v, N_0(U^{i-1})\geq m-b_v, \\
X_i, & \mbox{otherwise},
\end{array}\right.
\end{eqnarray*}
where $N_x(U^{i-1})$ is the number of times the symbol $x$ occurs in
the sequence $U^{i-1}$. Define a random variable $V$ by $V:=v$ if
and only if $S\in [a_v,b_v]$. Define a random variable $K$ by
$K:=\min \{i|i \in [1,m], N_1(U^i)\geq a_V, N_0(U^{i})\geq m-b_V
\}$. In other words, we define $U_1=X_1$, $U_2=X_2$, and so on,
until after $K$ steps, $U^K=X^K$ contains at least $a_V$ ones and at
least $(m-b_V)$ zeros, so that for arbitrary values of the remaining
sources $x_{K+1}^m$, $S$ definitely belongs to $[a_V, b_V]$, which
means that the desired function is determined. After the $K$-th
step, no information is sent, because $U_{K+1}^m=\mathbf 0$. One can
verify that $U^t$ satisfy the Markov chains and the conditional
entropy equality in (\ref{eqn:rateregion}). Intuitively speaking,
the above definition of $U^t$ corresponds to the following one-round
coding scheme: For each sample, the source nodes keep sending the
original data until there exists $v\in[1,v_{\max}]$ such that $a_v$
ones and $(m-b_v)$ zeros have appeared. Once it happens, the sum of
sources definitely falls into an $f'$-monochromatic interval
$[a_v,b_v]$ so that the desired function is determined. Thus the
computation for this sample is stopped. $K$ is the stopping time of
sending data.

For any source sequence $x^m$, the corresponding values $k$ and
$u^m$ satisfy $u^k=x^k$ and $u_{k+1}^m=\mathbf 0$. Therefore
\[p_{U^m}(u^m)= p_{U^k}(u^k) p_{U_{k+1}^m|U^k}(\mathbf
0|u^k)=p_{X^k}(x^k).\] Then we have
\begin{eqnarray}
R_{sum,1} &\leq& I(X^m;U^m)\nonumber\\&=& H(U^m)\nonumber\\ &=&
\eE\left(\log \frac{1}{p_{U^m}(U^m)}\right)\nonumber\\&=&
\eE\left(\log \frac{1}{p_{X^K}(X^K)}\right)\nonumber\\&=&
\eE(H(X^K|K))\nonumber\\&=& \eE(K) h(p)\nonumber\\&=&
h(p)\sum_{v=1}^{v_{\max}} p_V(v) \eE(K|V=v).\label{eqn:upperbound1}
\end{eqnarray}

We need to bound $\eE(K|V=v)$ with respect to the joint distribution
of $(K,V)$, which is given by: if $k\in [1,m], v\in[1,v_{\max}]$,
$p_{K V}(k,v)=\pP(X^k$ contain
 $a_v$ ones and $(m-b_v)$ zeros, but $X^{k-1}$ do
not $)$; otherwise $p_{K V}(k,v)=0$.

For each $v$, we can define another random variable $K_v'$ as the
number of iid Bernoulli$(p)$ trials to get $a_v$ ones and $(m-b_v)$
zeros. In other words, $p_{K_v'}(k)=\pP(Y^k$ contain
 $a_v$ ones and $(m-b_v)$ zeros, but $Y^{k-1}$ do
not $)$, where $Y_i\sim$ iid Bernoulli$(p)$ for $i\in \nN$. Note
that unlike $K$, which does not exceed $m$, $K_v'$ could be
arbitrarily large.

Since $\forall v\in[1,v_{\max}], \forall k \in [1,m],
p_{K_v'}(k)=p_{K V}(k,v)$, the conditional distributions of
$(K|V=v)$ and $(K_v'|K_v'\leq m)$ are the same. Therefore
\begin{equation}
\eE(K|V=v)=\eE(K_v'|K_v'\leq m)\leq
\eE(K_v').\label{eqn:upperbound2}
\end{equation}
The last step is because for any random variable $X$ and $\forall
a\in \rR$, $\eE(X|X\leq a)\leq \eE(X)$.

Then, define two independent random variables $W_{v,1}$ and
$W_{v,0}$ as follows: $W_{v,1}$ is the number of iid Bernoulli$(p)$
trials to get $a_v$ ones. $W_{v,0}$ is the number of iid
Bernoulli$(p)$ trials to get $(m-b_v)$ zeros. They are negative
binomial distributed random variables and $\eE(W_{v,1})=a_v/p$ and
$\eE(W_{v,0})=(m-b_v)/(1-p)$. Since if $Y^{W_{v,1}}$ contains $a_v$
ones and $Y_{W_{v,1}+1}^{W_{v,1}+W_{v,0}}$ contains $(m-b_v)$ zeros,
then $Y^{W_{v,1}+W_{v,0}}$ contains at least $a_v$ ones and
$(m-b_v)$ zeros, we have $K_v'\leq (W_{v,1}+W_{v,0})$, which implies
that
\begin{equation}\eE(K_v')\leq
\eE(W_{v,1})+\eE(W_{v,0})=a_v/p+(m-b_v)/(1-p).\label{eqn:upperbound3}\end{equation}
Combining (\ref{eqn:upperbound1}), (\ref{eqn:upperbound2}) and
(\ref{eqn:upperbound3}) leads to the statement of
Theorem~\ref{thm:lowersymmfuncBerp}(iii).


\footnotesize

\bibliography{newbibfile}
\end{document}